# Sparse inversion for derivative of log determinant

Shengxin Zhu[1][0000-0002-6616-6244] and Andrew J Wathen[2][0000-0001-9992-5588]

[1] Xi'an Jiaotong-Liverpool University, Suzhou 215123, P.R. China
`Shengxin.Zhu@xjtlu.edu.cn`
[2] University of Oxford, OX26GG, Oxford, UK
`wathen@maths.ox.ac.uk`

**Abstract.** Learning algorithms for Gaussian process, marginal likelihood methods or restricted maximum likelihood methods often require derivatives of log determinant terms. These log determinants are usually parameterized with variance parameters of the underlying statistical models. The first derivative of log determinant with many variance parameters for large data sets is usually computational prohibitive. This paper demonstrates that, when the underlying matrix is sparse, how to take the advantage of sparse inversion (selected inversion which share the same sparsity as the original matrix) to accelerate evaluating the derivative of log determinant.

## 1    Introduction

Many statistical models and machine learning algorithms often result in an optimization problem of a complicated target function involving log determinant terms. We take the linear mixed model as an example. Linear mixed model is a widely used supervised learning technique for knowledge discovery. They are widely used for various unbalanced, repeated, or missing data analysis, such as breeding, genome-wide association studies [30][42], recommendation systems [15] and knowledge discovery [7, 43]. The model is conceptually simple:

$$y = X\tau + Zu + e, \tag{1}$$

where $y$ is a vector of $n$ observable measurements or response variable, $e$ is the vector of model residuals, $\tau$ are fixed effects and $u \in R^{b \times 1}$ are random effects. X consists of $p$ columns of dummy variables which correspond to the so-called fixed effects. Z consists $b$ dummy columns which correspond to $b$ number of cross-memberships or clusters; each block is a random effect. $u$ and $e$ are unobservable random variables such that $E(u) = 0$, and $E(e) = 0$. $u$ and $e$ are assumed to be multivariate normally distributed with

$$var \begin{bmatrix} u \\ e \end{bmatrix} = \begin{bmatrix} G(\gamma) & 0 \\ 0 & R(\phi) \end{bmatrix}, \tag{2}$$

where the matrices G and R are parameterized with γ and φ respectively. Here γ and φ are vector parameters. For example, in the case the residuals e can be divided into b groups, the residuals are uncorrelated to each other. Only residuals between the intra-blocks share the same variance parameters, the variance matrix $R(\phi)$ is a direct sum of block diagonal matrices, say, $R(\phi) = \bigoplus_i^b \phi_i^2 I$. The parameter σ is a variance parameter which is referred to as the overall scale parameter. It follows that $y \sim N(X\tau, \sigma^2 H)$ where

$$\text{var}(y) = \text{var}(e) + Z\text{var}(u)Z^T = \sigma^2(R + ZGZ^T) := \sigma^2 H. \tag{3}$$

When the variance parameter $\kappa = (\sigma, \gamma, \phi)$ are known, the *best linear unbiased estimators* (BLUEs) for fixed effects $\hat{\tau}$, and the *best linear unbiased prediction* (BLUP) for the random effects $\tilde{u}$, satisfy the following *compacted* mixed model equations (MME)[22][38][36]

$$C \begin{bmatrix} \hat{\tau} \\ \tilde{u} \end{bmatrix} = \begin{bmatrix} X^T R^{-1} X & X^T R^{-1} Z \\ Z^T R^{-1} X & Z^T R^{-1} Z + G^{-1} \end{bmatrix} \begin{bmatrix} \hat{\tau} \\ \tilde{u} \end{bmatrix} = \begin{bmatrix} X^T R^{-1} y \\ Z^T R^{-1} y \end{bmatrix}. \tag{4}$$

Unbiased refers to the expectation of the estimation of $\hat{\tau}$ and $\tilde{u}$ satisfy $E(\hat{\tau}) = \tau$ and $E(\tilde{u}) = u$, where τ and u are unknown true values. Compacted refers that the dimension of the MMEs $p + b$, is smaller than the dimension of H which is $n \times n$ (before having derived the MMEs, to get the BLUEs and BLUP involves inverting H, see [22][38]). Further, [23] provides quantification of the standard errors of the fixed and random effects. The variance matrix of the estimate $\hat{\tau}$ and the prediction error $\hat{u} - u$ satisfy

$$\text{var} \begin{bmatrix} \hat{\tau} \\ \tilde{u} - u \end{bmatrix} = \sigma^2 C^{-1}. \tag{5}$$

It is clear now $C^{-1}$ is a variance-covariance matrix and its diagonal elements give variances $\hat{\tau}$ and $\tilde{u} - u$.

Problems become much more difficult but more realistic when the variance parameters κ are unknown. In this case, an iterative procedure has to be used to estimate the variance parameter κ. The restricted maximum likelihood (REML) is often preferred to estimate variance parameters [19][35]. Basically, one has to use an iterative algorithm to maximize the following restricted maximum log likelihood [18][20][24]

$$\ell_R = -\frac{1}{2}\{(n-p)\log \sigma^2 + \log \det H + \log \det(X^T H^{-1} X) + y^T P y / \sigma^2\} \tag{6}$$

$$= -\frac{1}{2}\{(n-p)\log \sigma^2 + \log \det C + \log \det R + \log \det G + y^T P y / \sigma^2\}, \tag{7}$$

where $H = R + ZGZ^T$,
   $P = H^{-1} - H^{-1}X(X^T H^{-1} X)^{-1} X^T H^{-1} = R^{-1} - R^{-1} W C^{-1} W^T R^{-1}$,
and $W = [X, Z]$. According to some basic Calculus results, we know that the stationary points of $\ell_R(\kappa)$ is one of the mostly likely point at which $\ell_R$ can realize the maximum value. Thus the variance parameter estimation problem is reduced to find the stationary points of $\ell_R$, say, $\partial \ell_R / \partial \kappa_i = 0$. It can be shown that the two formulas (6) and (7) are equivalent. See for example [43] for a derivation. As mentioned that the dimension of H is usually far larger than that of C. R and G usually enjoy simply structure. Therefore we shall use the formula (7) as our target function. The derivative of $\ell_R$, though complicated, can be derived through standard matrix differentials procedures [21, p.309, eq. 8.6]. For example according to (7), one have to compute the derivatives of log determinant terms like

$$\frac{\partial \log \det C}{\partial \kappa_i} = tr\left(C^{-1}\frac{\partial C}{\partial \kappa_i}\right).$$  (8)

Such terms are essential for the Newton method and the Fisher-scoring algorithms.

In contrast to evaluating the likelihood itself which can be computed efficiently by modern sparse techniques [6,8,9,12,33,39], computing the derivatives of the log likelihood is more difficult. Because it is has been shown that the inverse of an irreducible non-singular matrix is always full [13, Thm 6], this indicates that algorithms for computing all the elements of the inverse of an irreducible matrix takes $O(n^2)$ space and usually has $O(n^3)$ complexity, even the matrix is sparse. Such algorithms are not feasible for large scale problems like those arising from breeding and genome-wide association study (GWAS).

Luckily, a key observation is that when C is sparse, $\partial C/\partial \kappa_i$ is usually part of C. The calculation of the trace terms only uses elements of $C^{-1}$ corresponding to nonzero elements of C, which means the complete $C^{-1}$ is not necessary, and only those elements of $C^{-1}$ at the position of nonzero of C are necessary. We shall refer to this part as the sparse inverse of C. Thanks to the Takahashi-Fagan-Chin formulae [14], fast compute the sparse inverse is possible when the Cholesky factorization or $LDL^T$ factorization is available. Such a formula has been efficiently implemented for general sparse matrices with LU factorization in the name of inverse multi-frontal methods [4,5] and for the symmetric matrices in the name of selected inversion algorithm [29]. It should be pointed out that the Takahashi-Fagan-Chin formulae have been successfully applied in breeding for decades [34]; recent applications involving the selected inversion algorithm includes electronic structure calculations [28], computing Green's function [26], and astronomy [3] and so on. The presentation here is based on partial of our knowledge transfer project [40].

This technical note focuses on efficient methods to factorize the large sparse matrices C and economical ways to get its sparse inverse. Most of the presented results are not fundamentally new, but they are scattered in several places. One practical implementation needs to integrate as many as these scattered results. With the emergence of large scale statistical computing which receives increasingly attention, it is very likely that more people will be interested such integrated technical details which are closely related to computing the derivatives of the log likelihood. Our techniques by introducing the underlying principles of related techniques, and supply additional background and most relevant references.

The remaining of this paper is organized as follows. Section 2 introduces an efficient factorization algorithm for symmetric positive definite matrices, the $LDL^T$ factorization, followed by two approaches to derive the selected inversion algorithm in Section 3. Section 4 discusses the fill-in reducing ordering and symbolic analysis to improve the performance of the factorization and inversion. Numerical examples are presented in Section 5.

## 2    Factorization

Factorizing a symmetric positive definite matrix $C$ is usually done by the Cholesky factorization. This can be efficiently implemented by the $C = LDL^T$ factorization, where $L$ is a unit lower triangular matrix with diagonal elements 1. $D$ is a diagonal matrix. The classical Cholesky factorization $C = \tilde{L}\,\tilde{L}^T$ can be obtained form the $LDL^T$ factors by setting $\tilde{L} = LD^{1/2}$ ; in practice, such a transform is not necessary and the matrix $D$ can be stored in the diagonal part of the matrix $L$ to save memory.

The $LDL^T$ factorization algorithm can be derived from the following well known formula which can be used to derive the inversion formula.

$$C = \begin{bmatrix} \alpha & a^T \\ a & \hat{C} \end{bmatrix} = \begin{bmatrix} 1 & \\ \ell & I \end{bmatrix} \begin{bmatrix} \alpha & \\ & S \end{bmatrix} \begin{bmatrix} 1 & \ell^T \\ & I \end{bmatrix} \tag{9}$$

$$= \begin{bmatrix} 1 & \\ \ell & \hat{L} \end{bmatrix} \begin{bmatrix} \alpha & \\ & \hat{D} \end{bmatrix} \begin{bmatrix} 1 & \ell^T \\ & \hat{L}^T \end{bmatrix} \tag{10}$$

where $\ell = a/\alpha$, $S = \hat{L}\hat{D}\hat{L}^T = \hat{C} - aa^{T}/\alpha = \hat{C} - \ell a^T$. $S$ is usually referred to as the *Schur complements*. By recursively using the formula (9) n-1 times to these Schur complements generated in the process, one can derive one algorithm for the $LDL^T$ factorization. Efficient implementation of the $LDL^T$ factorization algorithm can be found in [8].

Suppose there are $m_1$ elements in the first column of $L$, this includes the first diagonal element of $D$ which can be saved in the diagonal position of $L$, then computing $\ell$ in (9) requires $m_1 - 1$ floating point operations (FLOPs); with consideration of the symmetry, updating the Schur complement $S$ requires $m_1(m_1 - 1)$ FLOPs. Let $m_i$ $i = 1,2,\cdots,n$ ($n = p + b$ is the total number of fixed and random effects in the linear mixed model) be the number of non-zero elements in the $i$th column of $L$. Then computing the $L$ and $D$ factors for the $LDL^T$ factorization requires

$$\sum_{i=1}^{n} m_i^2 - n \tag{11}$$

floating point operations.

At moment, little can be said on $m_i$, but it happens to be that the $m_i$ equals to the number of nonzero elements in first column of $C$; the non-zero pattern of the $i$th column of $L$ is the same as that of the lower triangular part of $i$th column of C is not the usual case. Taking the second column of L as an example, it has the same sparse pattern as the first column of $S = \hat{C} - aa^T/\alpha$; the nonzero elements in S, neglecting the numerical cancellations, are those in positions corresponding to non-zero elements in $\hat{C}$ and those in positions corresponding to non-zero elements in the out product $aa^T$. Those non-zero elements in aa^T which locates the positions of zero elements in C are called *fill-in.* The worst case for the formula (9) and (10) arises when the first column of C is dense; in such a case, the out product $aa^T$ is dense and it results in a dense Schur component S, thus one can not get benefit of the sparsity of C. Techniques for reordering the matrix C to reduce as many as fill-in aim at making the first columns as sparse as possible, further discussion is presented later.

With the LDL$^T$ factorization, the evaluation of the logdet C is readily:

$$\text{logdet } C = \sum_{i=1}^{p+b} \log d_{ii}, \tag{12}$$

where $d_{ii}$ is the $i$th diagonal element of D. This can be used to compute the log-likelihood, which is useful for inference and model selection.

## 3    Selected inversion

There are two ways to derive inversion formula based on the LDL$^T$ factorization. The first way is given by Takahashi, Fagan and Chin [14] for general non-singular matrices. We illustrate the idea for symmetric positive definite matrices. Suppose A has an LDL$^T$ factorization, then $Z = A^{-1} = L^{-T} D^{-1} L^{-1}$ and thus $ZL = L^{-T} D^{-1}$, add Z to both side of $ZL = L^{-T} D^{-1}$, one can get the Takahashi-Fan-Chin formula for symmetric positive definite matrices:

$$Z = L^{-T} D^{-1} + Z(I - L). \tag{13}$$

Notices the first part $L^{-T} D^{-1}$ is an upper triangular matrix, thus only $Z(I - L)$ contributes to the strictly lower triangular part of Z. The elements of Z can be represented as

$$Z_{ij} = \begin{cases} [Z(I-L)]_{ij} = \sum_{k>j} Z_{ik} L_{kj} = \sum_{k>j} Z_{ki} L_{kj}, & \text{for } i > j, \\ [L^{-T} D^{-1} + Z(I-L)]_{ii} = d_{ii}^{-1} + [Z(I-L)]_{ii}, & for \ i = j. \end{cases} \tag{14}$$

Let $\tilde{z}_j$ be the $j$th column of the strictly lower triangular part of Z, $\ell_j$ be the $j$th column of the strictly lower triangular part of L, then $\tilde{z}_j = \tilde{Z} \ell_j$, where $\tilde{Z}$ is the sub-matrix of Z corresponding to the rows of $\tilde{z}_j$. In such a way, computing those elements in $\tilde{z}_j$ in the positions of non-zero elements in $\ell_j$ only requires those elements in Z corresponding to rows and columns corresponding to non-zero rows of $\ell_j$. It is clear that $Z_{nn} = d_{nn}^{-1}$, and the computations can be carried on from the bottom–right corner to the up-left corner of the inverse Z.

Lin et al. use another approach to derive the inversion formula more clearly [29]. They use the inverse of the $2 \times 2$ block matrix C as partitioned in (9):

$$C^{-1} = \begin{bmatrix} \alpha^{-1} + \ell^T S^{-1} \ell & -\ell^T S^{-1} \\ -S^{-1} \ell & S^{-1} \end{bmatrix}. \tag{15}$$

This is the fundamental formula for the selected inversion algorithm. Suppose S is the $1 \times 1$ matrix which is generated in the final step of the LDL$^T$ factorization in (9) and (10). Then $S^{-1} = d_{nn}^{-1}$. Now suppose the inverse of the Schur component of order $m \times m$, $m > 1$ has been computed. The inverse of the Schur component of order $(m + 1) \times (m + 1)$ can be computed in a partion like (15). Such a process is equivalent to the Takahashi-Fan-Chin formula: to compute the $j$th column of the lower triangular part of the inverse, only the matrix of $\bar{Z}$ is needed, $\bar{Z}$ is equivalent to the inverse of the Schur component $S^{-1}$.

Computing those elements in $-S^{-1}\ell$ in the positions corresponding to non-zero elements in $\ell$ takes $2(m_i - 1)^2 + m_i - 1$ FLOPs (with consideration of the minus sign), where $m_i$ is defined as before, the number of non-zero elements in $\ell$. Updating the diagonal entries in the (1,1) block requires $2(m_i - 1)$ FLOPs. In the previous step, only a subset of $S^{-1}$ is computed, precisely, the subset in positions corresponding to non-zeros elements of $LDL^T$factors of $S$. One may ask whether this subset is enough for next step, computing the subset of $S^{-1}\ell$? Some results on *elimination tree* give a positive answer, see [29][31] and reference therein for details.

With consideration of the symmetry of $C$, obtaining the subset of the inverse of $C^{-1}$ requires

$$2 \sum_i^{n-1} (m_i - 1)^2 + 3 \sum_i^{n-1} (m_i - 1) = 2(\sum_{i=1}^n m_i^2 - n) - (\sum_{i=1}^n m_i - n) \quad (16)$$

FLOPs (one has to consider $m_n = 1$ for the simplification). The computation amount in (16) is twice of that in (11) minus the number of non-zero elements in the strictly lower triangular part of $L$.

One should pay heed to that such FLOPs counts do not include non-numerical operations in the omitted index matching procedures, which features the FIND algorithm, see [25, 26, 27]. Careful implementation of the FIND algorithm can bring benefit.

Such a selected inversion algorithm computes a subset of elements of $C^{-1}$, the subset contains all the elements in $C^{-1}$ in the positions corresponding to non-zero elements of $C$. These subset of elements are ready for evaluating the derivative of logdet$C$ in (8).

## 4    Reordering and symbolic analysis

Both (11) and (16) show that computations amount depends on the number of non-zero elements in L. The number of non-zero elements in the $i$th column of L, $m_i$, consists of two parts, say, $m_i = m_{C_i} + m_{f_i}$, where $m_{C_i}$ is the number of elements in points where in C there are non-zeros elements, and $m_{f_i}$is the number of fill-in. Neglecting the numerical cancellation, $m_i \geq m_{C_i}$. These $m_{f_i}$ fill-in are not ultimate necessary to evaluate the derivatives; and reducing such fill-in can reduce computations in the factorizing, triangular solves and the selected inverting process. That's why reordering the equations is needed.

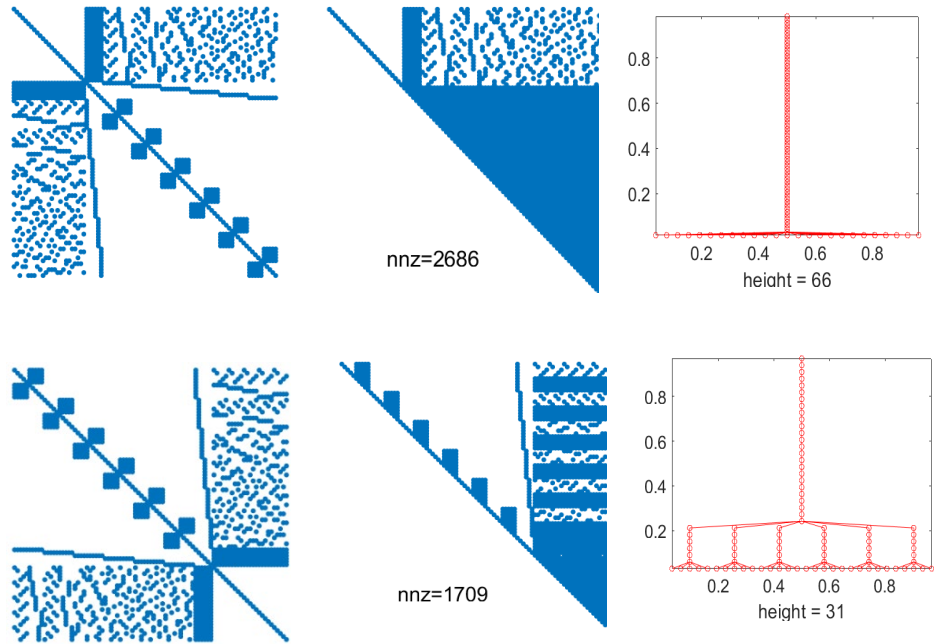

**Fig. 1.** The sparsity of matrix C, its Cholesky factor and elimination tree. First row for the original matrix, second row for reordered matrix.

An ordering corresponds to a permutation (matrix or vector) P, which changes the rows or columns of a matrix. Given a matrix C, finding permutation (matrices) P and Q such that the number of non-zeros in the factorization PCQ is minimized is the *fill-minimization problem*; when C is symmetric, it requires $Q = P^T$. Obtaining a good solution to the reordering problem plays an important role in designing sparse direct solvers. The problem is an NP-hard problem [32][41]. Therefore, heuristics are employed. Commonly used methods are minimum degree ordering [17] and nested dissection ordering [16], the latter is more favored for large problem from 2D and 3D spatial discretization and usually is coupled with minimum degree ordering. The fill-reducing ordering algorithm adopted here is the approximate minimum degree ordering [1, 2], in Matlab this corresponds to the amd function. It can collaborate with the fill-reducing software package METIES[47].

Reordering is a precursor to factorization. Together with reordering, *symbolic analysis/factorization* can predict the structure of L before *numerical* factorization starts. Symbolic factorization works on the sparse pattern rather than numerical values, and thus it saves a lot of memory and can be processed efficiently. The non-zero pattern of *L* and column counts $m_i$ can be analyzed by *elimination tree* [9, p.56][31] which is also used to analyze the data flow dependence. The factorization starts with a leaf nodes and ends with the root node. The branches describe the data dependence. A shorter and wider tree enjoys more parallelism, see Fig.1 for illustration. Once famil-iar with the elimination tree, one can realize that reordering is not only for reducing fill-in, but also for shortening and broadening an elimination tree to reduce the sequential operations steps and to increase parallelism.

In Matlab, the symbolic analysis can be achieved by the function symbfact (though this is always an implicit step of many software). It returns the sparse pattern and the column counts of L. With the column counts $m_i$, one can pre-allocate memory space for L before numerical factorization starts, such a way can avoid the dynamical memory allocation for the fill-in during the numerical factorization. Compared with dynamical memory allocation, pre-allocation save more overheads and thus more efficient.

## 5 Numerical Examples

First presented are variety trial problems using the linear mixed model. And then we apply the techniques to general positive definite matrices. Since the performance of sparse matrix computation are largely limited by the memory bandwidth of the underlying computing platform, we investigate the relationship between the size of L and the timing results in Fig. 2.

### 5.1 Variety trials benchmark problem

These examples are based on a second-stage analysis of a set of variety trials, i.e. based on variety predicted values from each trial. Trials are conducted in a number of years across a number of locations (centers). See Table 1.

The generating program allows one to specify the number of years, total number of centers and proportion of centers used per year, the number of control varieties (used every year), the number of test varieties entering the system per year and the average persistence of the test varieties, and the proportion of missing varieties per trial, where proportions of things are selected, They are sampled at random, and the life of each variety is generated from a Poisson distribution. This gives a three-way crossed structure (year*variety*site) with some imbalance. In the current model, all terms except a grand mean are fitted as random. The random terms are generated as independent and identically distributed normal distribution with variance components generated from a test program with similar structure used for the original SAS REML program, so it is just a variance components model.

They are sampled at random, and the life of each variety is generated from a Poisson distribution. This gives a three-way crossed structure (year*variety*site) with some imbalance. In the current model, all terms except a grand mean are fitted as random. The random terms are generated as independent and identically distributed normal distribution with variance components generated from a test program with similar structure used for the original SAS REML program, so it is just a variance components model.

**Table 1.** Date sets for the benchmark problem

| DataSet | year | center | varity | y.c | y.v | v.c | units | v/y | y/v | c.v |
|---|---|---|---|---|---|---|---|---|---|---|
| Prob1 | 12 | 22 | 130 | 132 | 673 | 2518 | 6667 | 56.1 | 5.2 | 10 |
| Prob 2 | 15 | 25 | 160 | 180 | 888 | 3527 | 9595 | 59.2 | 5.6 | 10 |
| Prob 3 | 22 | 25 | 188 | 264 | 1177 | 4215 | 12718 | 53.5 | 6.3 | 12 |
| Prob 4 | 25 | 25 | 262 | 300 | 1612 | 5907 | 17420 | 64.5 | 6.2 | 12 |
| Prob 5 | 25 | 25 | 390 | 300 | 2345 | 8625 | 25334 | 93.8 | 6.0 | 15 |
| Prob 6 | 25 | 35 | 390 | 425 | 2345 | 12249 | 35887 | 93.8 | 6.0 | 15 |
| Prob 7 | 30 | 35 | 470 | 510 | 3013 | 15087 | 46113 | 100.4 | 6.4 | 20 |
| Prob 8 | 30 | 35 | 620 | 510 | 3835 | 19737 | 58685 | 127.8 | 6.2 | 20 |
| Prob 9 | 35 | 40 | 720 | 700 | 4522 | 26432 | 81396 | 129.2 | 6.3 | 20 |
| Prob 10 | 40 | 50 | 820 | 1000 | 5262 | 37701 | 118403 | 131.6 | 6.4 | 20 |

**Table 2.** Symbolic analysis of the $LDL^T$ factorization and selected inversion

| Prob | No Effects | C | | | L | | | FLOPs count | |
|---|---|---|---|---|---|---|---|---|---|
| | | nnz | $n_z$ | ‰ | nnz | $n_z$ | ‰ | $LDL^T$ | selinv |
| Prob 1 | 3488 | 56946 | 16.3 | 9.4 | 112618 | 32.3 | 18.5 | 8943842 | 1778554 |
| Prob 2 | 4796 | 80946 | 16.9 | 7.0 | 172023 | 35.9 | 15.0 | 17175555 | 34183883 |
| Prob 3 | 5892 | 105059 | 17.8 | 6.1 | 273315 | 46.4 | 15.7 | 40768817 | 81270211 |
| Prob 4 | 8132 | 144240 | 17.7 | 4.4 | 377761 | 46.5 | 11.4 | 60714709 | 121059789 |
| Prob 5 | 11711 | 209235 | 17.9 | 3.1 | 507711 | 43.4 | 7.4 | 75897428 | 151298856 |
| Prob 6 | 15470 | 291318 | 18.8 | 2.4 | 718701 | 46.5 | 6.0 | 149074099 | 297444967 |
| Prob 7 | 19146 | 370799 | 19.4 | 2.0 | 1020414 | 53.3 | 5.6 | 270835518 | 540669768 |
| Prob 8 | 24768 | 473891 | 19.1 | 1.5 | 1196903 | 48.3 | 3.9 | 290699965 | 58027795 |
| Prob 9 | 32450 | 648237 | 20.0 | 1.2 | 1779662 | 54.8 | 3.4 | 6000925570 | 1200103928 |
| Prob 10 | 44874 | 932054 | 20.8 | 0.9 | 2817463 | 62.8 | 2.8 | 1391099157 | 2779425725 |

**Table 3.** Timing results for the benchmark problem

| Prob | reordering | $LDL^T$ | selinv | LDU | T-F-C |
|---|---|---|---|---|---|
| Prob1 | 0.0121 | 0.093 | 0.0111 | 0.0405 | 0.0327 |
| Prob 2 | 0.0197 | 0.0127 | 0.0187 | 0.0633 | 0.0754 |
| Prob 3 | 0.0266 | 0.0173 | 0.0389 | 0.0944 | 0.1268 |
| Prob 4 | 0.0433 | 0.0244 | 0.0581 | 0.1361 | 0.1988 |
| Prob 5 | 0.0633 | 0.0440 | 0.0762 | 0.2171 | 0.2732 |
| Prob 6 | 0.0854 | 0.0505 | 0.1540 | 0.2769 | 0.5056 |
| Prob 7 | 0.1177 | 0.0679 | 0.2402 | 0.3796 | 0.8944 |
| Prob 8 | 0.1741 | 0.0854 | 0.2686 | 0.4968 | 1.0049 |
| Prob 9 | 0.2507 | 0.1280 | 0.5361 | 0.7336 | 1.9238 |
| Prob 10 | 0.4010 | 0.8054 | 1.2112 | 1.3276 | 4.2305 |

## 5.2 More general examples

For general problems, we can consider using general symmetric positive definite matrices as covariance matrices. Various problems form University of Florid sparse matric collection are used in Table 4 and Table 5 to show the efficiency of the method.

**Table 4.** Symbolic analysis and selected inversion for various problems

| Problems | No. Effects | C | | | L | | | FLOPs count | |
|---|---|---|---|---|---|---|---|---|---|
| | | nnz | $n_z$ | ‰ | nnz | $n_z$ | ‰ | $LDL^T$ | selinv |
| bcsstk14 | 1806 | 32630 | 18.07 | 20. | 109078 | 60.40 | 66.8 | 9137434 | 18167596 |
| bcsstk28 | 4410 | 111717 | 25.33 | 11.5 | 340375 | 77.18 | 35.0 | 35033411 | 69730857 |
| bcsstk38 | 8032 | 181746 | 22.63 | 5.6 | 736620 | 91.71 | 22.8 | 118966988 | 237205388 |
| bcsstk18 | 11948 | 80519 | 6.74 | 1.1 | 637035 | 53.32 | 8.9 | 130092475 | 259559863 |
| cbuckle | 13681 | 345098 | 25.22 | 3.7 | 2354457 | 172.10 | 25.2 | 629317072 | 1256293368 |
| bodyy4 | 17546 | 69548 | 3.96 | 0.5 | 572607 | 32.63 | 3.7 | 5686867 | 102618673 |
| bodyy5 | 18589 | 73721 | 3.97 | 0.4 | 636443 | 34.24 | 3.7 | 66255816 | 123893778 |
| bodyy6 | 19366 | 76787 | 3.97 | 0.4 | 687941 | 35.52 | 3.7 | 69991661 | 139324747 |
| raefsky4 | 19779 | 668284 | 33.79 | 3.4 | 7304797 | 369.32 | 37.3 | 5307655672 | 10608046326 |
| bcsstk36 | 23052 | 583096 | 25.29 | 2.2 | 2732511 | 118.54 | 10.3 | 622744081 | 1242778703 |
| crystm03 | 24696 | 304233 | 12.32 | 1.0 | 5037495 | 203.9 | 16.5 | 1975947597 | 3946887385 |
| wathen | 36441 | 301101 | 8.26 | 0.5 | 1845324 | 50.64 | 2.8 | 246966843 | 492124803 |
| thermal1 | 82654 | 328566 | 3.98 | 0.1 | 2659592 | 32.18 | 0.8 | 292072446 | 581567954 |
| shipsec1 | 140874 | 1854525 | 13.16 | 0.2 | 41000460 | 291.04 | 4.1 | 52300864586 | 104560865586 |
| bmwca 1 | 148770 | 5395186 | 36.27 | 0.5 | 90924001 | 611.17 | 8.2 | 114866116347 | 229644457463 |
| pwtk | 217918 | 5877775 | 26.94 | 0.2 | 60054622 | 275.58 | 2.5 | 46821917900 | 93583999096 |
| Parabolic fem | 525825 | 2100225 | 3.99 | 0.0 | 36132647 | 68.72 | 0.3 | 17110241646 | 34184876470 |
| ecology2 | 999999 | 2997995 | 3.0 | 0.0 | 43752523 | 45.75 | 0.1 | 20047008630 | 40049264736 |
| G3 circuit | 1585478 | 4623152 | 2.92 | 0.0 | 189102138 | 119.27 | 0.2 | 299948175092 | 599708833574 |

**Table 5.** Timing results for various problems form University of Florida

| Prob | reorder-ing | $LDL^T$ | selinv | LDU | T-F-C |
|---|---|---|---|---|---|
| bcsstk14 | 0.0009 | 0.0066 | 0.0078 | 0.0194 | 0.0372 |
| bcsstk28 | 0.0017 | 0.0124 | 0.0293 | 0.0712 | 0.1152 |
| bcsstk38 | 0.0045 | 0.0300 | 0.0990 | 0.1487 | 0.3044 |
| bcsstk18 | 0.074 | 0.0246 | 0.1136 | 0.1482 | 0.4091 |
| cbuckle | 0.0088 | 0.0743 | 0.5180 | 0.4660 | 1.7940 |
| bodyy4 | 0.0059 | 0.0305 | 0.0474 | 0.1097 | 0.1960 |
| bodyy5 | 0.0059 | 0.0313 | 0.0556 | 0.1251 | 0.2158 |
| bodyy6 | 0.0061 | 0.0330 | 0.0629 | 0.1364 | 0.2444 |
| raefsky4 | 0.0147 | 0.2338 | 4.6353 | 1.3349 | 11.5883 |
| bcsstk36 | 0.0108 | 0.0954 | 0.5170 | 0.5379 | 1.8475 |
| crystm03 | 0.0137 | 0.1399 | 1.5345 | 0.7527 | 3.3403 |
| wathen | 0.0095 | 0.0645 | 0.2095 | 0.4088 | 1.1996 |
| thermal1 | 0.0501 | 0.1096 | 0.2645 | 0.5538 | 1.0480 |
| shipsec1 | 0.2250 | 1.4231 | 47.5587 | 8.8262 | 134.2235 |
| bmwca 1 | 0.1618 | 3.2910 | 107.9326 | 111.6025 | 428.1022 |
| pwtk | 0.1640 | 2.6615 | 40.3312 | 13.0838 | 131.9626 |
| Parabolic | 0.4778 | 1.9287 | 16.1115 | 7.2807 | 48.4882 |
| ecology2 | 0.4012 | 1.7227 | 17.2147 | 9.9427 | 75.0454 |
| G3 circuit | 1.5675 | 8.8011 | 309.9106 | ------- | ------- |

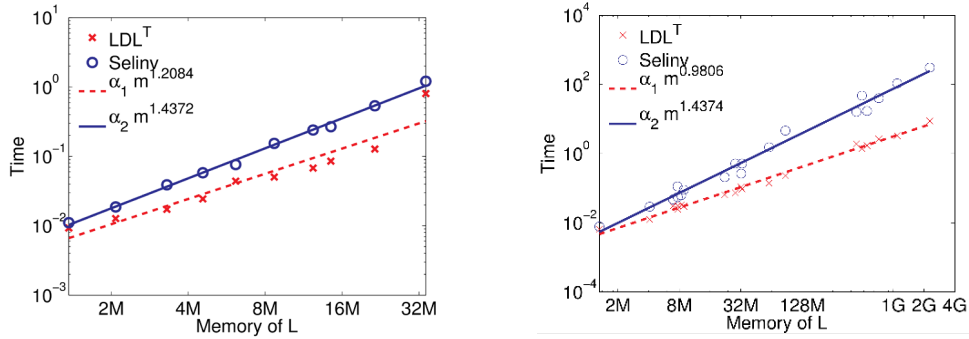

**Fig. 2.** Relationship between time and the memory size of L, Table 3(left) Table 4(right).

## 6 Discussion and conclusion

The log determinant problems frequently appears in many machine learning algorithms. While modern learning algorithms more and more relay on hardware and software packages. This note shows that neat algorithms can significantly improve the evaluation of the first derivative of log determinant terms. Derivative methods for REML and any other log likelihood related methods were believed as computationally intensive, recent development in hardware and software make it possible for large scale problems. In particularly, the sparse factorization methods and selected inversion techniques can reduce the computation in 1 or 2 orders. While the whole procedure of REML involves more involved mathematics. More sophisticated techniques like auto differentiation can be used to evaluate the derivatives during the factorization and inverting process. And the average information techniques can accelerate the approximation of Jacobian matrix in the Newton iteration. The reader is directed to [37] and [45] for details. When the matrix C is dense, there are recently scalable methods to calculate the log determinants terms [48][49] and related software package [50]. As cited in these publications, such methods for dense matrix is based on the fast decaying of the eigenvalues of the Gaussian kernel function [51].

## Acknowledgements

The authors would like to thank the reviewers for their constructive feedback to improve this paper. The research is supported by Natural Science Foundation of China (NSFC.11501044), Jiangsu Science and Technology Basic Research Program (BK20171237), Key Program Special Fund of XJTLU (KSF-E-21, KSF-E-32), Research Development Fund of XJTLU (RDF-2017-02-23), Research Enhance Fund of XJTLU (REF-18-01-04) and partially supported by NSFC (No.11771002, 11571047, 11671049, 11671051, 6162003, and 11871339).